\documentclass[prl,aps,twocolumn,showpacs,superscriptaddress]{revtex4-1}
\usepackage{hyperref}
\hypersetup{colorlinks=true, citecolor=blue, filecolor= black, linkcolor= blue, urlcolor= blue}
\usepackage{graphicx}
\usepackage{dcolumn}
\usepackage{bm}
\usepackage{float}
\begin{document}


\title{Two-dimensional boron on Pb $\mathbf{(110)}$ surface}

\author{Xin-Ling He}
\affiliation{Key Laboratory of Weak-Light Nonlinear Photonics and School of Physics, Nankai University, Tianjin 300071, China}

\author{Xiao-Ji Weng}
\affiliation{Key Laboratory of Weak-Light Nonlinear Photonics and School of Physics, Nankai University, Tianjin 300071, China}

\author{Yue Zhang}
\affiliation{Key Laboratory of Weak-Light Nonlinear Photonics and School of Physics, Nankai University, Tianjin 300071, China}

\author{Zhisheng Zhao}
\affiliation{State Key Laboratory of Metastable Materials Science and Technology, Yanshan University, Qinhuangdao 066004, China}

\author{Zhenhai Wang}
\affiliation{Moscow Institute of Physics and Technology, Dolgoprudny, Moscow Region 141700, Russia}
\affiliation{Department of Geosciences, Center for Materials by Design, and Institute for Advanced Computational Science, Stony Brook University, Stony Brook, New York 11794, USA}
\affiliation{College of Telecommunication and Information Engineering, Nanjing University of Posts and Telecommunications, Nanjing, Jiangsu 210003, China}

\author{Bo Xu}
\affiliation{State Key Laboratory of Metastable Materials Science and Technology, Yanshan University, Qinhuangdao 066004, China}

\author{Artem R. Oganov}
\affiliation{Moscow Institute of Physics and Technology, Dolgoprudny, Moscow Region 141700, Russia}
\affiliation{Department of Geosciences, Center for Materials by Design, and Institute for Advanced Computational Science, Stony Brook University, Stony Brook, New York 11794, USA}
\affiliation{Skolkovo Institute of Science and Technology, 3 Nobel St., Moscow 143026, Russia}
\affiliation{School of Materials Science, Northwestern Polytechnical University, Xi'an 710072, China}

\author{Yongjun Tian}
\affiliation{State Key Laboratory of Metastable Materials Science and Technology, Yanshan University, Qinhuangdao 066004, China}

\author{Xiang-Feng Zhou}
\email{xfzhou@nankai.edu.cn}
\affiliation{Key Laboratory of Weak-Light Nonlinear Photonics and School of Physics, Nankai University, Tianjin 300071, China}

\author{Hui-Tian Wang}
\affiliation{Key Laboratory of Weak-Light Nonlinear Photonics and School of Physics, Nankai University, Tianjin 300071, China}
\affiliation{National Laboratory of Solid State Microstructures and Collaborative Innovation Center of Advanced Microstructures, Nanjing University, Nanjing 210093, China}

\begin{abstract}
\noindent We simulate boron on Pb $(110)$ surface by using \textit{ab initio} evolutionary methodology. Interestingly, the two-dimensional (2D) Dirac $Pmmn$ boron can be formed because of good lattice matching. Unexpectedly, by increasing the thickness of 2D boron, a three-bonded graphene-like structure ($P2_{1}/c$ boron) was revealed to possess double anisotropic Dirac cones. It is 20 meV/atom lower in energy than the $Pmmn$ structure, indicating the most stable 2D boron with particular Dirac cones. The puckered structure of $P2_{1}/c$ boron results in the peculiar Dirac cones, as well as substantial mechanical anisotropy. The calculated Young's modulus is 320 GPa$\cdot$nm along zigzag direction, which is comparable with graphene.
\end{abstract}

\pacs{62.23.Kn, 62.20.de, 73.20.At}


\maketitle
Graphene is featured by the Dirac cone in the band structure, which leads to the fractional quantum Hall effect, ultrahigh carrier mobility, and some other novel properties \cite{R01,R02,R03,R04}. The fancy properties of graphene inspire search for other 2D materials with Dirac cones, and at least seven 2D carbon allotropes with Dirac cones were predicted, i.e., $\alpha$, $\beta$, $\delta$, 6,6,12--, 14,14,14--, 14,14,18-graphyne, and phagraphene \cite{R05,R06,R07,R08}. Boron has a short covalent radius and the flexibility to adopt $sp^{2}$ hybridization as carbon, which favors the formation of low-dimensional forms (clusters, nanotubes, fullerenes and so on) \cite{R09,R10,R11,R12,R13,R14,R15,R16,R17,R18,R19,R20,R21}. Owing to the rich variety of bonding configurations, ranging from the common two-center two-electron bonds to the rare eight-center two-electron bonds, an extreme structural diversity is anticipated in 2D boron. Up to now, several boron-related phases were predicted to have distorted Dirac cones, but this prediction remains to be realized \cite{R14,R17,R21}. Most recently, borophenes were successfully grown on Ag $(111)$ surface under ultra-high vacuum conditions \cite{R19,R20}. These monolayer boron sheets with triangular lattice (without vacancy) or 1/6 vacancies were synthesized and confirmed to be metallic. However, growing multilayer boron sheets on Ag $(111)$ substrate is difficult due to the weak boron-silver interaction. When the boron coverage exceeds one monolayer on Ag $(111)$, 3D boron clusters tend to form easily \cite{R19,R20}. In contrast, the boron-copper interaction is stronger than boron-silver interaction, thereby 2D multilayer boron sheets were synthesized on copper foils by using chemical vapor deposition and turned out to be semiconducting with a bandgap of 2.35 eV \cite{R18}. Hunting for 2D boron with Dirac cones are still a big challenge for both experimentalists and theoreticians \cite{R22,R23}. Whether $Pmmn$ boron and $P6/mmm$ boron can be synthesized and whether other 2D Dirac semimetallic boron allotropes can be formed on specific substrates are therefore unknown. In this work, the face-centered cubic Pb metal was selected as an alternative substrate to grow $Pmmn$ boron because the Pb $(110)$ surface matches the lattice of $Pmmn$ boron very well. Moreover, Pb is more reactive than Cu and Ag, thus the B-Pb interaction may be strong enough to grow multilayer boron sheets.

\begin{figure}[t]
\begin{center}
\includegraphics[width=8cm]{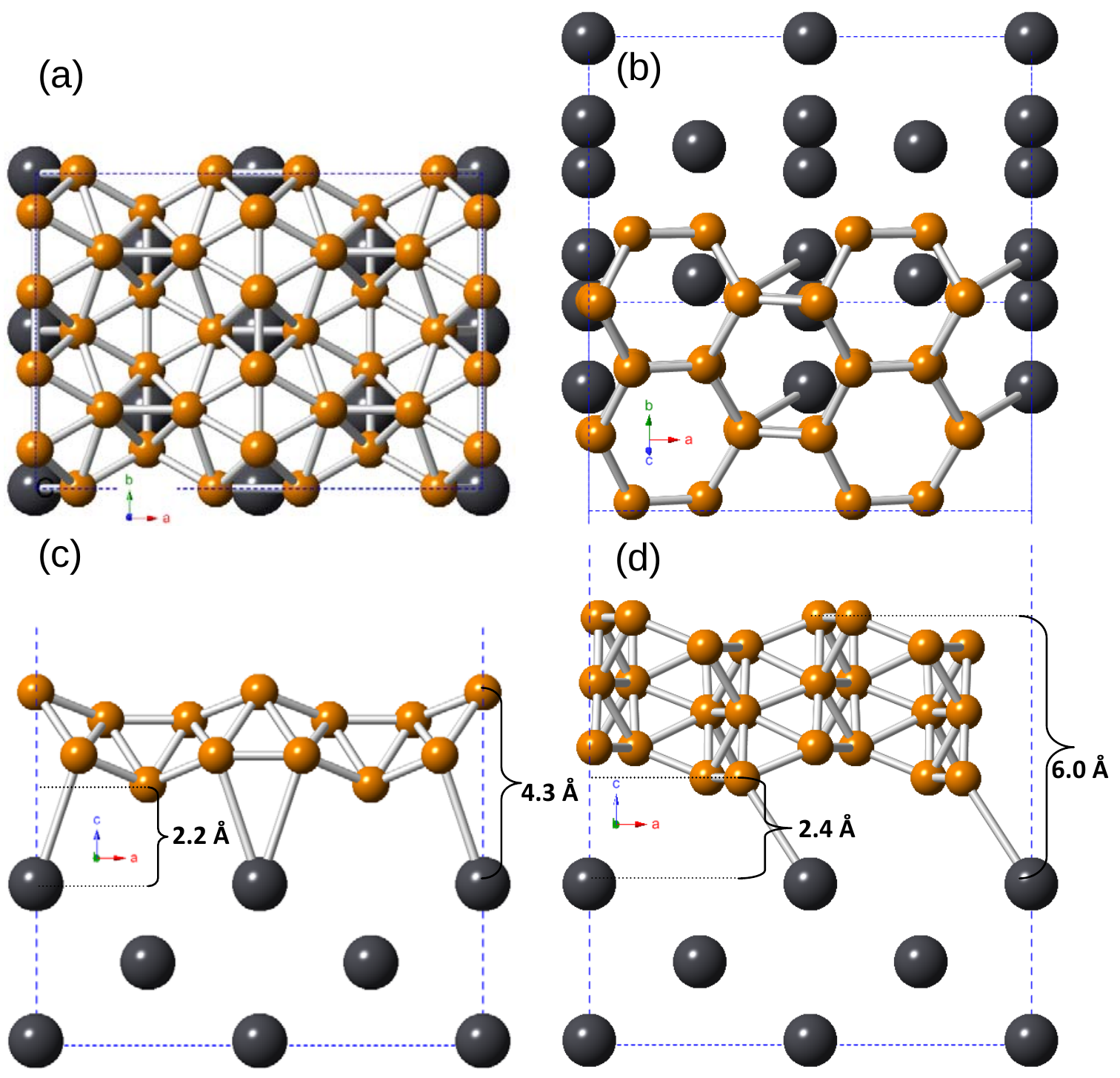}
\caption{%
(Color online) Structures of 2D boron on Pb $(110)$ substrate. [(a) and (c)] Projection of $Pmmn$ boron along [001] and [010] directions, [(b) and (d)] Projection of $P2_{1}/c$ boron along [001] (downward) and [010] directions. The B and Pb atoms are colored in orange and dark gray, respectively.}
\end{center}
\end{figure}

\begin{figure}[h]
\begin{center}
\includegraphics[width=8cm]{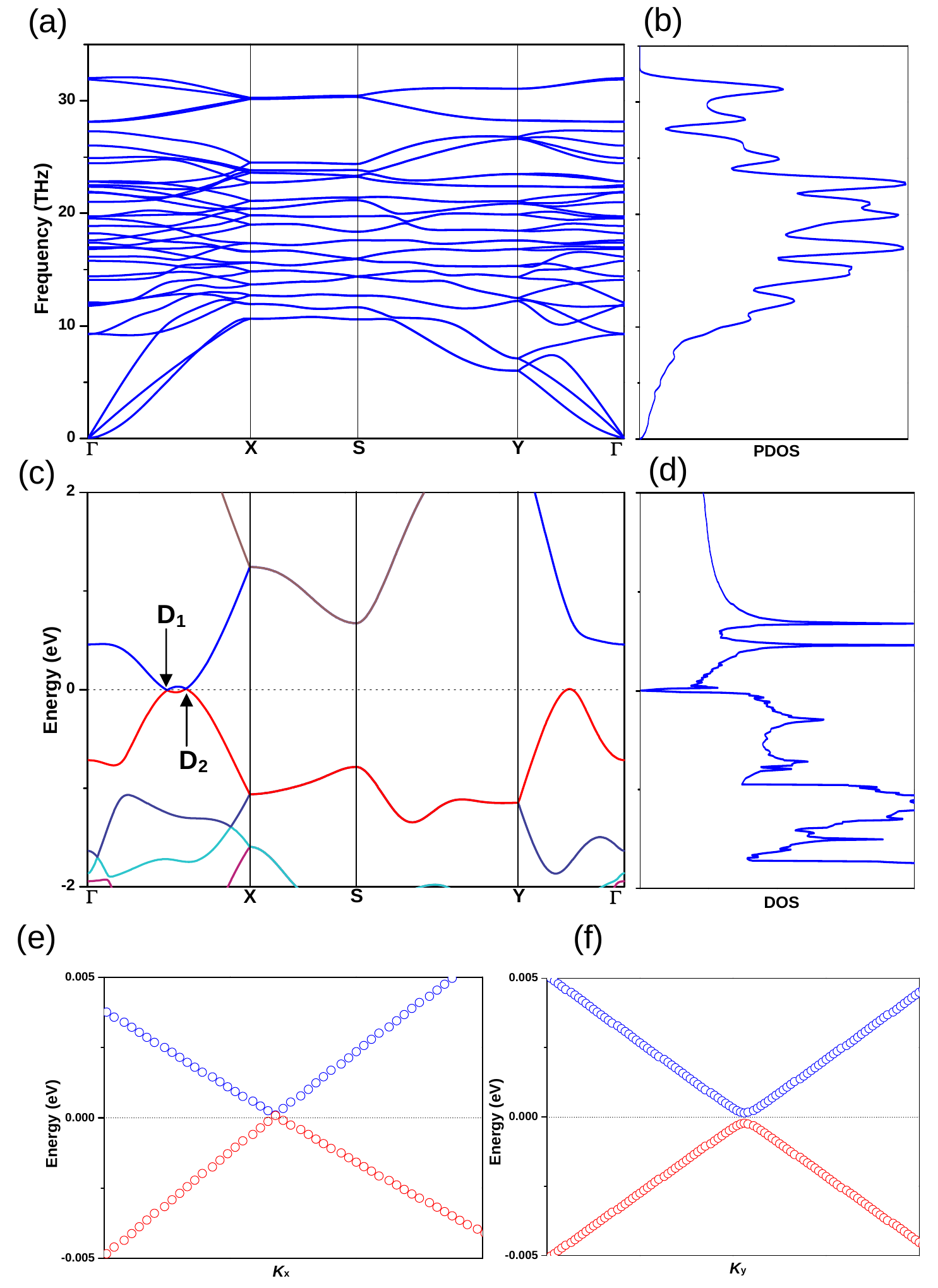}
\caption{%
(Color online) (a) Phonon dispersion of freestanding $P2_{1}/c$ boron at ambient conditions, (b) Phonon density of states (PDOS), (c) Band structure of freestanding $P2_{1}/c$ boron. The crossing bands meet at D$_{1}$ and D$_{2}$ points, (d) Density of states (DOS), (e) and (f) Band structure in the vicinity of D$_{1}$ point along $k_{x}$ and $k_{y}$ directions, respectively.}
\end{center}
\end{figure}

The evolutionary structure searches were conducted with 5, 6, 7, 8, 10, 12, 14, and 16 boron atoms per unit cell using the \textsc{uspex} code \cite{R24,R25,R26}, which has been successfully applied to various surface systems \cite{R27,R28,R29}. The vacuum layer, surface layer, and substrate geometry are pre-specified. The initial thickness of the surface and vacuum layers were set to 4~{\AA} and 15~{\AA}, and were allowed to change during relaxation. The structure relaxations used the all-electron-projector augmented wave method \cite{R30} as implemented in the Vienna \textit{ab initio} simulation package (VASP) \cite{R31}. The exchange-correlation energy was computed within the generalized gradient approximation (GGA) with the functional of Perdew, Burke, and Ernzerhof (PBE) \cite{R32}. The plane wave cutoff energy of 450 eV and the uniform $\Gamma$-centered $k$-points grids with resolution of $2 \pi \times 0.04$~\AA$^{-1}$ were used. Phonon dispersion curves were calculated by the supercell method using the \textsc{phonopy} package \cite{R33}.

\begin{table*}
\caption{Calculated lattice constants, atomic positions, and the total energy ($E_t$) of the $P2_{1}/c$ boron from GGA (PBE) results; the corresponding values for $P6/mmm$, $C2/m$(from Ref. [17]), $\alpha$-sheet, $Pmmn$, $Pmmm$, and $\alpha$-boron (from Ref. [14]) are also listed for comparison.}
\begin{tabular*}{15cm}{@{\extracolsep{\fill}}cllllc}
\hline\hline
Phase & $a$    & $b$    & $c$    & Atomic positions & $E_t$ \\
      &({\AA}) &({\AA}) &({\AA}) &                  & (eV/atom) \\
\hline
$P6/mmm$$^{\rm 17}$         & 2.87 & 2.87 & 15.0  & B1 (0.0 0.0 0.443)     & -6.27 \\
                            &      &      &       & B2 (0.333 0.667 0.382) &  \\
$\alpha$-sheet$^{\rm 14}$   & 5.07 & 5.07 & 13.00 & B1 (0.667 0.333 0.5)   & -6.28 \\
                            &      &      &       & B2 (0.667 0.667 0.5)   &  \\
$Pmmn$$^{\rm 14}$           & 4.52 & 3.26 & 13.00 & B1 (0.5 0.753 0.584)   & -6.33 \\
                            &      &      &       & B2 (0.185 0.5 0.531)   &  \\
$C2/m$$^{\rm 17}$           & 5.68 & 2.82 & 15.00 & B1 (0.808 0.5 0.409)   & -6.34 \\
                            &      &      &       & B2 (0.006 0.0 0.397)   &  \\
                            &      &      &       & B3 (0.15 0.0 0.510)    &  \\
$P2_{1}/c$                  & 3.18 & 4.85 & 15.00 & B1 (0.301 0.173 0.624) & -6.35 \\
                            &      &      &       & B2 (0.559 0.329 0.522) &  \\
                            &      &      &       & B3 (0.299 0.334 0.422) &  \\
$Pmmm$$^{\rm 14}$           & 2.88 & 3.26 & 13.00 & B1 (0.0 0.243 0.657)   & -6.36 \\
                            &      &      &       & B2 (0.5 0.5 0.621)     &  \\
                            &      &      &       & B3 (0.5 0.0 0.567)     &  \\
$\alpha$-boron$^{\rm 14}$   & 4.90 & 4.90 & 12.55 & B1 (0.803 0.197 0.976) & -6.68 \\
                            &      &      &       & B2 (0.119 0.238 0.892) &  \\
\hline\hline
\end{tabular*}
\end{table*}

The Pb $(110)$ substrate was constructed by rectangular lattice with the constants a=4.95~{\AA} and b=3.5~{\AA}, which are close to those of $Pmmn$ boron (a=4.52~{\AA} and b=3.26~{\AA}). With such substrate, the monolayer 2D boron was formed as containing six boron atoms per cell. This structure is composed of the buckled triangular B$_{3}$ units with distorted pentagonal and heptagonal holes (Fig. S1), which is completely different from those boron monolayers growing on Ag $(111)$ and Cu $(111)$ surfaces (the latter consists of B$_{3}$ units and contains hexagonal holes) \cite{R13,R15,R34}. The diversity of structural motifs can be attributed to different substrate-boron interactions. When adding seven boron atoms to the substrate, bilayer structures appeared first, looking very similar to $Pmmn$ boron with one vacancy located at the bottom boron chains. Therefore, by adding eight boron atoms to Pb $(110)$ surface, $Pmmn$ boron was exactly formed with a thickness of ~4.3~{\AA} away from the substrate. This structure is very stable even without any distortions (Figs. 1a and 1c), suggesting that $Pmmn$ boron is likely to be realized on Pb substrate. For comparison, calculations show that $Pmmn$ boron is unstable and transformed into a disordered-like phase on the rectangular Ag $(111)$ due to larger lattice mismatch and weak Ag-B interactions (Fig. S2) \cite{R34}, which may explain the absence of $Pmmn$ boron in recent experiments \cite{R19,R20}. In order to evaluate the thermodynamic stability for various 2D boron polymorphs, the formation energy is defined as $E_{f}$ = (1/$N$) ($E_{t}$ -- $E_{sub}$ -- $N$ $\times$ $E_{B}$), where $E_{t}$ and $E_{sub}$ are the total energies of the whole system and the substrate, respectively. $E_{B}$ is the energy per atom in bulk $\alpha$-boron and $N$ is the number of boron atoms \cite{R13}. The formation energy decreases with increasing $N$, which means that 2D structures become more stable with increasing thickness as they approach to the bulk state. It remains nearly constant ($\sim$0.38 eV/atom) when $N$ is greater than or equal to twelve atoms per unit cell (Fig. S3), revealing the high energy stability in 2D B$_{12}$ \cite{R34}. Freestanding relaxation of this structure shows that it has a monoclinic structure with $P2_{1}/c$ symmetry, which consists of three buckled graphene-like layers, mutually offset in the basal plane (Figs. 1b and 1d).  The lattice constants for the $P2_{1}/c$ boron are a=3.18~{\AA}, b=4.85~{\AA}, and $\beta$=$90.2^{\circ}$ (unique axis b), and three inequivalent atomic positions are B$_{1}$ (0.301 0.173 0.624), B$_{2}$ (0.559 0.329 0.522), and B$_{3}$ (0.299 0.334 0.422). The first and third graphene-like layers are built from the protruding B$_{1}$ and the sagging B$_{3}$ atoms, while the second layer is constructed by B$_{2}$ atoms. The total energy of $P2_{1}/c$ boron is 0.08 eV/atom, 0.07 eV/atom, 0.02eV/atom, and 0.01 eV/atom lower in energy than the $P6/mmm$, $\alpha$-sheet, $Pmmn$, $C2/m$ structures \cite{R14,R17}, but are 0.01 eV/atom and 0.33 eV/atom higher in energy than $Pmmm$ boron and bulk $\alpha$-boron \cite{R14}, showing that the $P2_{1}/c$ phase is metastable (Table 1). The phonon dispersion curve and phonon density of states (PDOS) establish that $P2_{1}/c$ boron is dynamically stable (Figs. 2a and 2b), indicating that it may be exfoliated from the Pb $(110)$ substrate.

The hexagonal graphitic boron monolayer is energetically unfavorable because electrons are insufficient to fill all of the bonding $\sigma$-- and $\pi$--orbitals, resulting in a metallic plane. Therefore, remarkable electronic properties of honeycomb sheet like those in graphene, such as Dirac cones, are usually absent in the graphitic 2D boron system \cite{R35}. Previously, non-graphene-like 2D boron ($Pmmn$ boron) was predicted to have a direction-dependent Dirac cone \cite{R14}. It was followed immediately by the proposal of 2D ionic boron composed of graphene-like plane and B$_{2}$ atom pairs, which possess double Dirac cones with massless Dirac fermions \cite{R17}. Here, three-bonded graphene-like layers in the $P2_{1}/c$ boron not only enhance the energetic stability, but are also the key factors for the emergence of peculiar Dirac cones. The band structure of $P2_{1}/c$ boron shows the valence and conduction bands meeting in two points (named as D$_{1}$ and D$_{2}$ points) along the $\Gamma$--X direction, while opening band gaps along $\Gamma$--Y, Y--S, and S--X directions (Fig. 2c). The zero density of states (DOS) at the Fermi level confirms that $P2_{1}/c$ boron is a semimetal (Fig. 2d). The highest valence and the lowest conduction bands calculated from GGA-PBE method shows that double particular Dirac cones are formed at the crossing points. These cones are little different from those of graphene, graphyne, or borophenes because two Dirac points stay close together \cite{R01,R05,R14,R36}, but still have perfect linear shapes (Figs. 2e and 2f). According to the time reversal symmetry, as shown in Fig. 3, these cones are located at the sides of the first Brillouin zone (BZ). The Fermi velocity was calculated as $v_{F}$ = (1/$\hbar$)$\partial$$E$/$\partial$$k$. The slope of crossing bands at D$_{1}$ point in the $k_x$ direction is 7 eV{\AA} ($v_{F}$ = $0.17 \times 10^6$ m/s) and --4.9 eV{\AA} ($v_{F}$ = $0.12 \times 10^6$ m/s). In the $k_y$ direction, the slopes of the bands are $\pm$5.1 eV{\AA}, equivalent to a Fermi velocity of $0.12 \times 10^6$ m/s. At D$_{2}$ point, the Fermi velocities are $0.17 \times 10^6$ m/s and --$0.13 \times 10^6$m/s in the $k_{x}$ direction, $\pm$$0.11 \times 10^6$ m/s in the $k_{y}$ direction. Therefore, the Fermi velocities at the crossing points are more or less the same, which means that the Dirac cones have similar size and arc sharp (Fig. 3c). The anisotropic Fermi velocities also imply the direction-dependent massless Dirac cones, but are lower than those of graphene ($0.82 \times 10^6$ m/s).
\begin{figure}[h]
\begin{center}
\includegraphics[width=8cm]{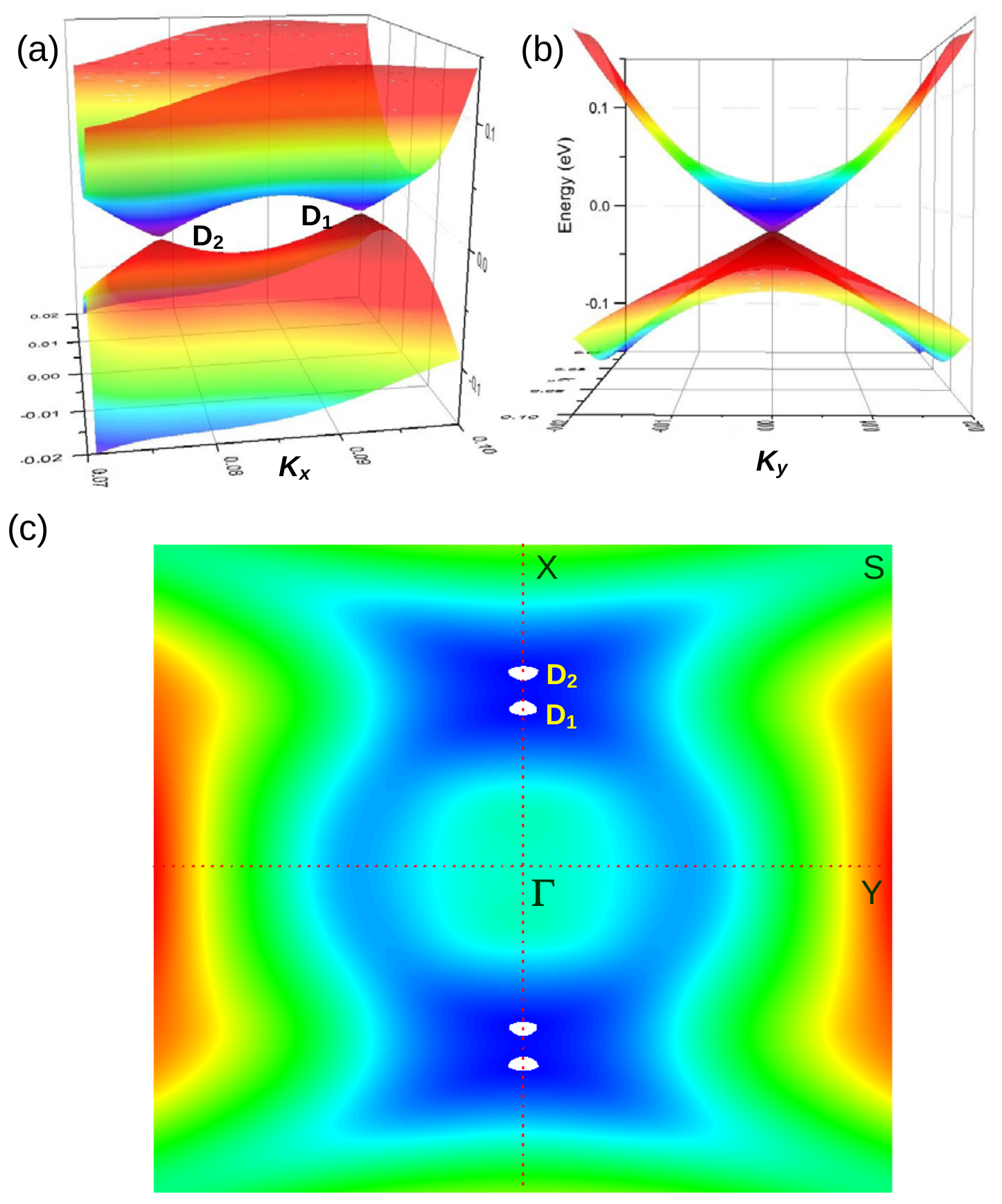}
\caption{%
(Color online) Dirac cones of $P2_{1}/c$ boron formed by the valence and conduction bands in the vicinity of D$_{1}$ and D$_{2}$ points. (a) Along $k_{x}$ direction, (b) Along $k_{y}$ direction, (c) The energy difference of the lowest conduction band and the highest valence band. The high symmetry points in the first Brillouin zone were labeled as $\Gamma$, $X$, $S$, and $Y$.}
\end{center}
\end{figure}

\begin{figure}[h]
\begin{center}
\includegraphics[width=8cm]{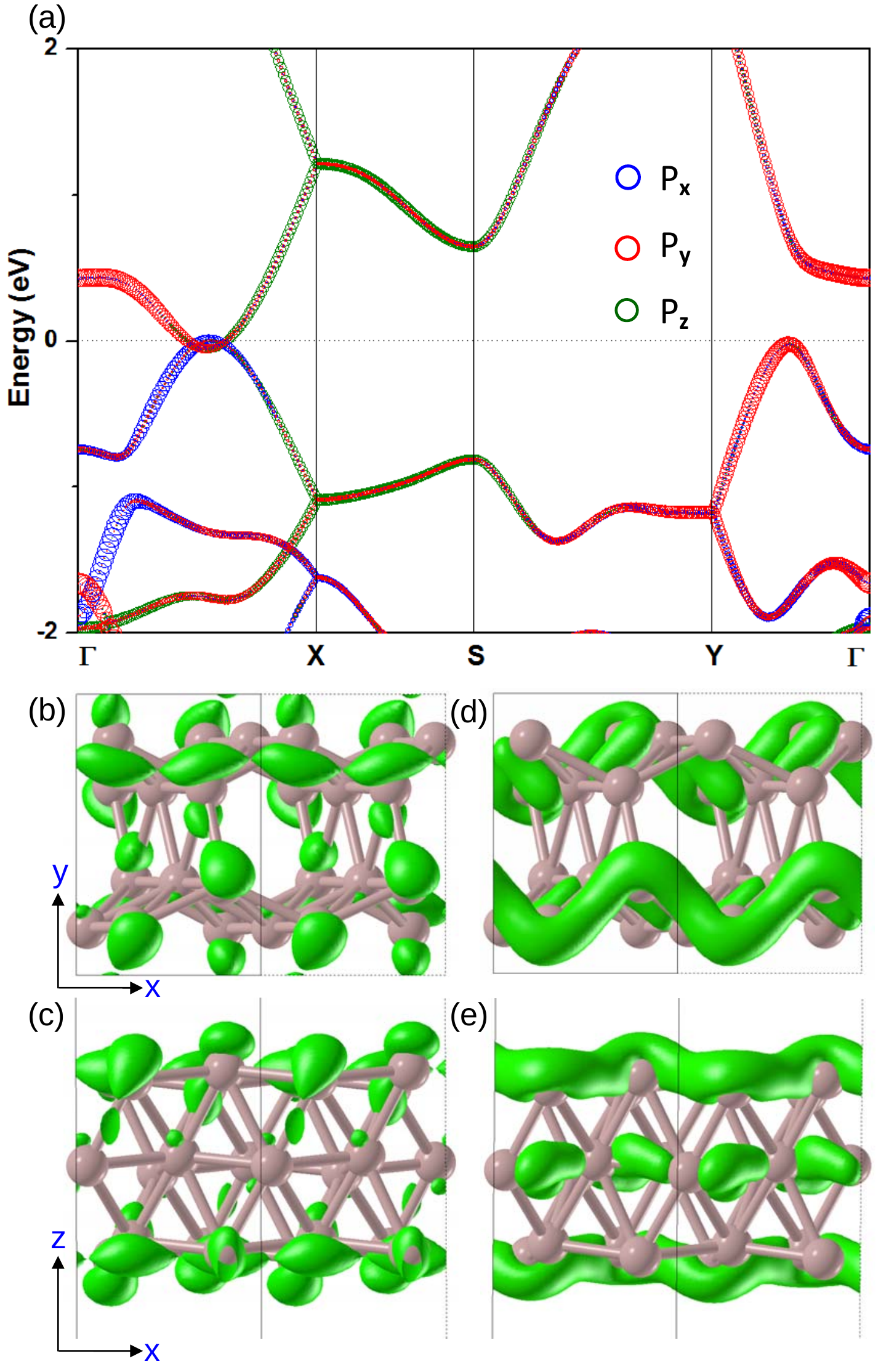}
\caption{%
(Color online) Orbital-resolved band structure and band-decomposed charge density of $P2_{1}/c$ boron at D$_{1}$ point. (a) Orbital-resolved band structure, [(b) and (c)] Projection of the charge density of the lowest conduction band along [001] and [100] directions, [(d) and (e)] Projection of the charge density of the highest valence band along [001] and [100] directions.}
\end{center}
\end{figure}
To explore the origin of the distorted Dirac cones, the orbital-resolved band structure of $P2_{1}/c$ boron is shown in Fig. 4a (to make the figure clearer, the $s$ orbital is ignored since its contribution is negligible). The contributions from different orbitals are illustrated by circles of different colors and sizes. Between D$_{1}$ and D$_{2}$ points above the Fermi level, the highest valence band is dominated by the $p_{x}$ orbital; below the Fermi level, the $p_{y}$ orbitals make major contribution to the lowest conduction band. Therefore, around the Fermi level, the electronic property is dominated by the in-plane states ($p_{x,y}$ orbitals), which also can be validated by the partial densities of states of three inequivalent atoms (see Fig. S4). At D$_{1}$ point, the primary contributions for the highest valence band and the lowest conduction band originate from the $p_{x}$ and $p_{y}$ orbitals accordingly; the $p_{z}$ orbitals play a secondary role. At D$_{2}$ point, the origin of the highest valence band is quite similar, whereas the $p_{z}$ orbitals play a major role in the formation of the lowest conduction band. Furthermore, the band-decomposed charge density of the lowest conduction band at D$_{1}$ point is dominantly distributed within the topmost and bottom layers (Figs. 4c and 4d). The in-plane states and  the out-of-plane states ($p_{z}$ orbitals) are mainly derived from the sagging B$_{3}$ atoms and protruding B$_{1}$ atoms, respectively. The charge density of the highest valence band is distributed among three layers along the $x$ direction. The $p_{x}$ orbitals originating from the protruding B$_{1}$ and middle layer B$_{2}$ atoms are arranged in buckling zigzag patterns, and are predominantly responsible for the charge density distribution (Figs. 4e and 4f). Although $P2_{1}/c$ boron has graphene-like structure, the origin of Dirac cones (dominated by the in-plane states) are different from those of graphene and graphynes (the crossing bands are derived from the $p_{z}$ orbitals exclusively) \cite{R05}, or $Pmmn$ boron and $P6/mmm$ boron (dominated by the $p_{z}$ orbitals) \cite{R14,R17}. In addition, according to symmetry, three different classes of 2D materials with Dirac cones exist. The first class is formed by graphene and other materials that have the Dirac cones located at high-symmetry points. The second class is represented by $\beta$-graphyne, $Pmmn$ boron, etc., where their cones are located along high-symmetry lines. Materials of the third class have Dirac cones at generic points in the BZ \cite{R39}. For time-reversal-symmetric systems with spinless fermions, $Miert$ and $Smith$ summarized some 2D Dirac materials with different space groups, uncovered formation of 2D Dirac materials by using the minimal two-band model, and concluded that the mirror symmetry plays a decisive role in the emergence of Dirac fermions for the first two class materials \cite{R39}. For $P2_{1}/c$ boron, double distorted Dirac cones are located along the $\Gamma$--X direction; thereby it belongs to the second class. However, owing to the absence of mirror symmetry, the Dirac cones in $P2_{1}/c$ boron can be understood from the glide-reflection symmetry, being similar to that of phosphorene. The few-layered phosphorenes can transform from a semiconductor to a band-inverted semimetal under electric field \cite{R38}. The striking double Dirac cones in $P2_{1}/c$ boron may be merged into a single cone by moving some atoms (Fig. S4). Such atomic movement brings about mirror symmetry in modified $P2_{1}/c$ boron (see supplementary materials for structural parameters) while preserving its glide-reflection symmetry. However, such modified structures are thermodynamically unstable, and revert to $P2_{1}/c$ boron again \cite{R34}.

\begin{figure}[t]
\begin{center}
\includegraphics[width=8cm]{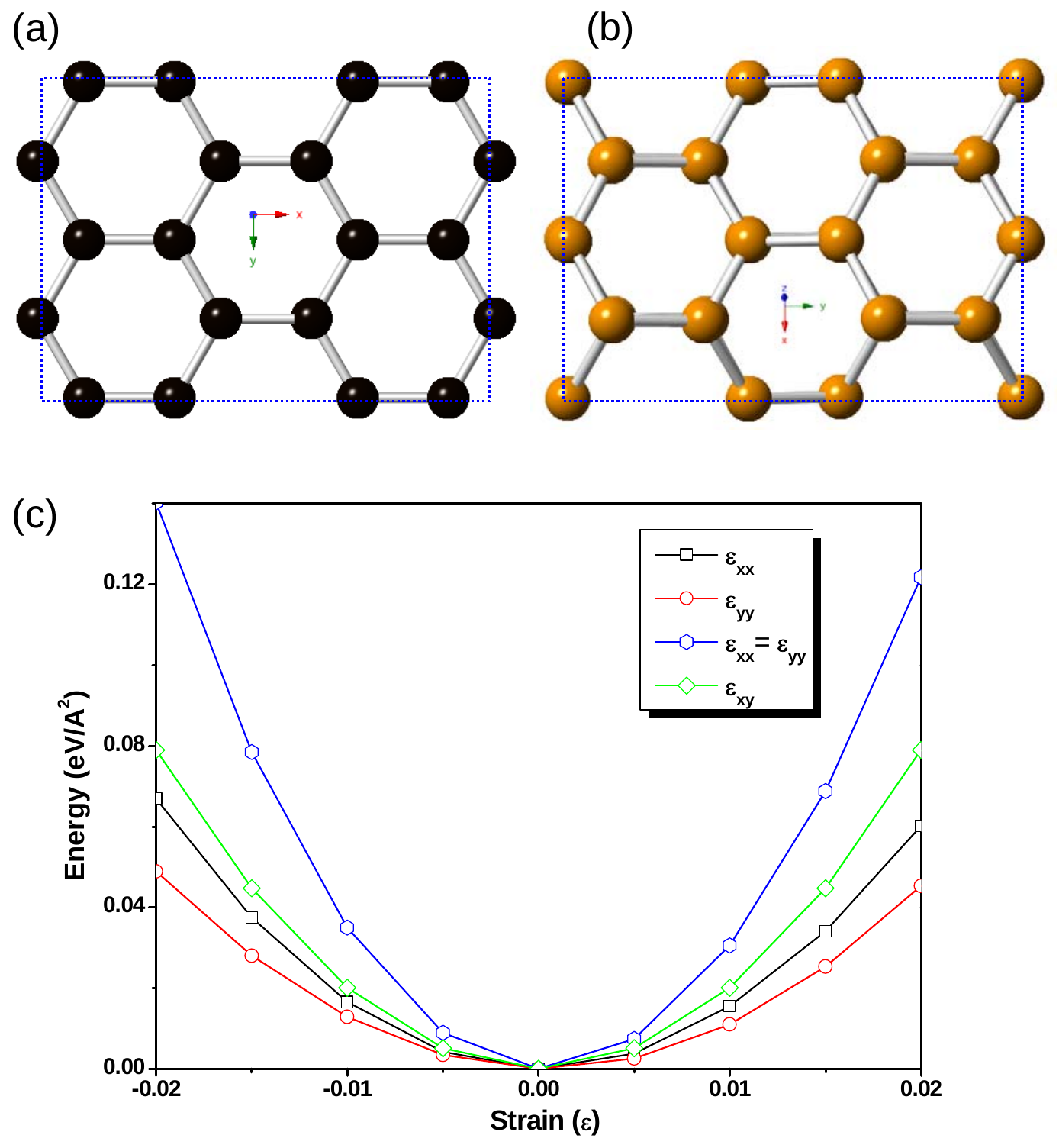}
\caption{%
(Color online) (a) Projection of the rectangular graphene structure along [001] direction. (b) Projection of the $2 \times 2 \times 1$ supercell of $P2_{1}/c$ boron along [101] direction. (c) Calculated energy vs. strain for the freestanding $P2_{1}/c$ boron.}
\end{center}
\end{figure}

$P2_{1}/c$ boron not only has graphene-like atomic structure, but also has peculiar electronic structure, such as particular Dirac cones. It is interesting to compare the mechanical properties of $P2_{1}/c$ boron and graphene. For 2D materials, using the standard Voigt notation \cite{R40}, the elastic strain energy per unit area can be expressed as U$_{\epsilon}$ = (1/2)C$_{11}$$\epsilon^2_{xx}$ + (1/2)C$_{22}$$\epsilon^2_{yy}$ + C$_{12}$$\epsilon_{xx}$$\epsilon_{yy}$ + 2C$_{66}$$\epsilon^2_{xy}$, where C$_{11}$, C$_{22}$, C$_{12}$, and C$_{66}$ are the elastic constants, corresponding to second partial derivatives of the energy with respect to strain. The in-plane Young's modulus can be derived from the elastic constants as E$_{x}$ = (C$_{11}$C$_{22}$--C$_{12}$C$_{21}$)/C$_{22}$ and E$_{y}$ = (C$_{11}$C$_{22}$--C$_{12}$C$_{21}$)/C$_{11}$. For graphene, in the same rectangular setting (Fig. 5a), the calculated C$_{11}$, C$_{22}$, C$_{12}$, and C$_{66}$ are 355 GPa$\cdot$nm, 352 GPa$\cdot$nm, 60 GPa$\cdot$nm, and 143 GPa$\cdot$nm, respectively. The Young¡¯s moduli are equal to ~ 345 GPa$\cdot$nm in both $x$ and $y$ directions. All of those are in excellent agreement with the reported theoretical (342 GPa$\cdot$nm) and experimental (340 GPa$\cdot$nm) values \cite{R03,R40}. In contrast, the calculated C$_{11}$, C$_{22}$, C$_{12}$, and C$_{66}$ for $P2_{1}/c$ boron are 331 GPa$\cdot$nm, 246 GPa$\cdot$nm, 53 GPa$\cdot$nm, and 103 GPa$\cdot$nm, respectively (Figs. 5b and 5c). The Young¡¯s modulus of $P2_{1}/c$ boron is equal to 320 GPa$\cdot$nm along the $x$ direction (zigzag direction), and 238 GPa$\cdot$nm along the $y$ direction (armchair direction). By checking the buckled structures carefully, strong covalent B-B bonds can be roughly divided into two classes: shorter B-B bonds ($\sim$1.76~{\AA} length) along zigzag direction, and longer ($\sim$1.82~{\AA}) bonds along the armchair direction. Thus, the shorter bonds result in higher Young¡¯s modulus.

In summary, two 2D semimetallic boron were predicted to grow on Pb $(110)$ substrate only by manipulating the surface thickness. Graphene-like boron sheet usually exists in metal borides, which are stabilized by the ionic interaction from metal atoms. For $P2_{1}/c$ boron, a three-bonded graphene-like boron structure was not only thermally and dynamically stable, but also had particular Dirac cones, providing better understanding for graphitic boron.

X. F. Z thanks Hongming Wen for valuable discussions. This work was supported by the National Science Foundation of China (Grant No. 11674176) and 111 Project (No. B07013). Z. W thanks the National Science Foundation of China (Grant No. 11604159) and the China Scholarship Council (No. 201408320093), the Natural Science Foundation of Jiangsu Province (Grant No. BK20130859). A. R. O. thanks the Government of Russian Federation (Grant No. 14.A12.31.0003), and the Foreign Talents Introduction and Academic Exchange Program (Grant No. B08040).



\end{document}